\documentclass{aa}  
\usepackage{graphicx}
\usepackage{txfonts}
\usepackage{subfigure}
\usepackage{inputenc}

\usepackage[colorlinks=true,citecolor=blue]{hyperref}

\begin{document}

\title{The circumstellar environment of EX~Lup: the SPHERE and SINFONI views\thanks{Based on observations collected at the European Organisation for Astronomical Research in the Southern Hemisphere under ESO programmes 089.C-0856(A) and 099.C-0147(B)} \\}

\author{E.\,Rigliaco\inst{1}
\and R.\,Gratton\inst{1} 
\and \'A. \,K\'osp\'al\inst{2,3,4}
\and D.\,Mesa\inst{1}
\and V.\,D'Orazi\inst{1}
\and P. \,\'Abrah\'am\inst{2,4}
\and S.\,Desidera\inst{1}
\and C.\,Ginski\inst{5,6} 
\and R.\,G. van Holstein\inst{6,7}
\and C.\,Dominik\inst{5}
\and A.\,Garufi\inst{8}
\and T.\,Henning\inst{9}
\and F.\,Menard\inst{10}
\and A.\,Zurlo\inst{11,12,13}
\and A.\,Baruffolo\inst{1} 
\and D.\,Maurel\inst{14}
\and P.\,Blanchard\inst{15}
\and L.\,Weber\inst{16}
}
\institute{
\inst{1}INAF/Osservatorio Astronomico di Padova, Vicolo dell'Osservatorio 5, 35122 Padova, Italy\\
\email{elisabetta.rigliaco@inaf.it}\\
\inst{2}Konkoly Observatory, Research Centre for Astronomy and Earth Sciences, Konkoly-Thege Mikl\'os \'ut 15-17, 1121 Budapest, Hungary\\
\inst{3}Max Planck Institute for Astronomy, K\"onigstuhl 17, D-69117 Heidelberg, Germany\\
\inst{4}ELTE E\"otv\"os Lor\'and University, Institute of Physics, P\'azm\'any P\'eter s\'et\'any 1/A, 1117 Budapest, Hungary\\
\inst{5}Anton Pannekoek Institute for Astronomy, University of Amsterdam, Science Park 904, 1098XH Amsterdam, The Netherlands\\
\inst{6}Leiden Observatory, Leiden University, PO Box 9513, 2300 RA Leiden, The Netherlands\\
\inst{7}European Southern Observatory, Alonso de C\'{o}rdova 3107, Casilla 19001, Vitacura, Santiago, Chile \\
\inst{8}INAF/Osservatorio Astrofisico di Arcetri, Largo Enrico Fermi 5, 50125 Firenze, Italy
\inst{9}Max Planck Institute for Astronomy, K\"onigstuhl 17, D-69117 Heidelberg, Germany\\
\inst{10}Univ. Grenoble Alpes, CNRS, IPAG, F-38000 Grenoble, France\\
\inst{11}N\'ucleo de Astronom\'ia, Facultad de Ingenier\'ia y Ciencias, Universidad Diego Portales, Av. Ejercito 441, Santiago, Chile \\
\inst{12}Escuela de Ingenier\'ia Industrial, Facultad de Ingenier\'ia y Ciencias, Universidad Diego Portales, Av. Ejercito 441, Santiago, Chile \\
\inst{13}Aix Marseille Universit\'e, CNRS, LAM (Laboratoire d'Astrophysique de Marseille) UMR 7326, 13388 Marseille, France \\
\inst{14}Univ. Grenoble Alpes, CNRS, IPAG, F-38000 Grenoble, France \\
\inst{15}Aix Marseille Universit\'e, CNRS, CNES,  LAM, Marseille, France\\ 
\inst{16}Geneva Observatory, University of Geneva, Chemin des Mailettes 51, 1290 Versoix, Switzerland\\
}

   \date{Received 04 May 2020 ; accepted 17 June 2020 }

 
  \abstract
 {EX~Lup is a well-studied T Tauri star that represents the prototype of young eruptive stars EXors. These are characterized by repetitive outbursts due to enhanced accretion from the circumstellar disk onto the star. In this paper we analyze new adaptive optics imaging and spectroscopic observations to study EX~Lup and its circumstellar environment in near-infrared in its quiescent phase. }
 {We aim at providing a comprehensive understanding of the circumstellar environment around EX~Lup in quiescence that builds upon the vast literature data. 
 }
 {We observed EX~Lup in quiescence with the high contrast imager SPHERE/IRDIS in the dual-beam polarimetric imaging mode to resolve the circumstellar environment in near-infrared scattered light. We complemented these data with earlier SINFONI spectroscopy, also taken in quiescence. 
 }
 {We resolve for the first time in scattered light a compact feature around EX~Lup azimuthally extending from $\sim$280$^{\circ}$ to $\sim$360$^{\circ}$, and radially extending from $\sim$0.3$^{\prime\prime}$ to $\sim$0.55$^{\prime\prime}$ in the plane of the disk. We explore two different scenarios for the detected emission. The first one accounts for the emission as coming from the brightened walls of the cavity excavated by the outflow whose presence was suggested by ALMA observations in the $J=3-2$ line of $^{12}$CO. The second one accounts for the emission as coming from an inclined disk. In this latter case we detect for the first time a more extended circumstellar disk in scattered light, which shows that a region between $\sim$10 and $\sim$30~au is depleted of $\mu$m-size grains. We compare the J$-$, H$-$ and K$-$band spectra obtained with SINFONI in quiescence with the spectra taken during the outburst, showing that all the emission lines were due to the episodic accretion event. 
  }
 {Based on the morphology analysis we favour the scenario in which the scattered light is coming from a circumstellar disk rather than the outflow around EX~Lup. We analyze the origin of the observed feature either as coming from a continuous circumstellar disk with a cavity, or from the illuminated wall of the outer disk or from a shadowed disk. 
 Moreover, we discuss what is the origin of the $\mu$m-size grains depleted region, exploring the possibility that a sub-stellar companion may be the cause of it. }


\titlerunning{EXLup}
\authorrunning{Rigliaco et al.}
  \maketitle
%

\section{Introduction}

EXors \citep{Herbig1989} are a class of pre-main sequence eruptive stars which show bursts of short duration (months–one year) with a recurrence time of years, showing accretion rates of the order of 10$^{-6}$ $-$ 10$^{-7}$~M$_{\odot}$/yr, and characterized by emission line spectra (e.g. \citealt{Herbig2008, Lorenzetti2009, Kospal2011, Sicilia2012, Antoniucci2013}). 
Studying this class of objects can provide useful insight into different open questions in star formation. 
As an example, many protostars are observed to be less luminous than theoretically expected (e.g., \citealt{Dunham2013}), and episodic accretion is  invoked as a solution for the luminosity problem. Understanding if all Sun-like young stars go through EXor-type outbursts at some point during their formation would clarify this aspect. Moreover, the physical mechanism causing the violent outbursts in EXors is still debated and EXors in quiescence are similar to regular Class II young stellar objects, with rather low-mass disks \citep{Liu2018}. Analyzing in more details this class of objects can shed light on the mechanisms powering the episodic outbursts. These objects provide us the possibility to study them both in quiescence and in outburst, and therefore we can observe the effect of episodic accretion on the circumstellar disk, in particular the planet-forming zone (e.g., the cristallization of silicate grains in EX Lup, \citealt{Abraham2009}).
For these reasons, it is indispensable to learn more about EXor circumstellar environments, and in particular about the circumstellar disk of EX Lup, the prototype of the class. 

EX Lup is a young ($\sim$2~Myr, \citealt{Garufi2018}) M0 T-Tauri star with M$_*$=0.6~M$_{\odot}$, 
at a distance of 157$\pm$0.9~pc (\citealt{Gaia2018, Bailer-Jones2018AJ}). It has exhibited its largest outburst in 2008 \citep{Jones2008}, brightening by about four magnitudes in visible light, and triggering a series of multi-wavelengths observations. Its most recent light curve is shown in \citet{Abraham2019}.   

Spectral energy distribution (SED) modeling of EX~Lup \citep{Sipos2009} evidences the presence of IR excess corresponding to a modestly flared dusty disk that extends from 0.2~au up to 150~au from the central star with a total mass of 0.025 M$_{\odot}$. This latter value is obtained by assuming that the dust model contained only amorphous silicates of olivine and of pyroxene types with a mass ratio of 2:1. Moreover, we note that in this work, 150 au was taken as a fixed outer boundary for the calculations and not fitted. The observations did not constrain the outer radius of the disk. 
The innermost regions of the disk around EX~Lup have been extensively studied \citep{Grosso2010, Kospal2014, Sicilia2015} and suggest infalling material reaching the star in hot spots via magnetospheric accretion. 
Based on hydrogen recombination lines, the mass accretion rate onto the star was $\sim$4$\times$10$^{-10}$~M$_{\odot}$/yr before the outburst in 2008 \citep{Sipos2009}, increased to 3$\times$10$^{-8}$--2$\times$10$^{-7}$~M$_{\odot}$/yr during the burst \citep{Sicilia2012}, and went back to $\sim$10$^{-10}$~M$_{\odot}$/yr afterwards \citep{Juhasz2012}.

The structure and dynamics of the outer disk has only recently started to be better studied. \citet{Kospal2016} detected with APEX $^{12}$CO(3--2), $^{12}$CO(4--3) and $^{13}$CO(3--2) molecular emission from the Keplerian disk around EX~Lup, but did not spatially resolve the disk. From the optically thin $^{13}$CO line they derived a total disk mass of 2.3$\times$10$^{-4}$~M$_{\odot}$, significantly lower than the one derived from the continuum data (assuming a canonical gas-to-dust ratio of 100) and SED-modeling. \citet{Hales2018} spatially resolved the dust emission from the disk in 1.3 mm continuum and a more extended Keplerian gas disk in $^{12}$CO(2--1), $^{13}$CO(2--1) and C$^{18}$O(2--1) using ALMA observations. Moreover, they pointed out a more extended and non-Keplerian $^{12}$CO(2--1) emission interpreted as the interaction of a molecular outflow with remnant ambient material. 
Comparing the EX~Lup $^{13}$CO/C$^{18}$O line ratios to those from the models of \citet{WilliamsBest2014}, \citet{Hales2018} obtained a total disk mass in gas of 5.4$\times$10$^{-4}$~M$_{\odot}$, and a total dust mass derived from the radiative transfer model fitting process of 1.0$\times$10$^{-4}$~M$_{\odot}$, yielding to a gas-to-dust ratio of 5.4. 
Gas-to-dust ratios lower than the typical interstellar medium value of 100 seem to be common around disks in Lupus \citep{Ansdell2016}. 

In the following we present new observations of EX Lup taken with the Very Large Telescope at ESO's Paranal Observatory in Chile. We employ polarimetric differential imaging (PDI) observations obtained with SPHERE/IRDIS in the H band to explore the circumstellar environment by tracing light scattered by the small ($\mu$m-sized) dust grains,  
and J-, H- and K bands medium-resolution near-infrared spectra acquired with SINFONI. 

The paper is organized as follows: in Sect.~2 we introduce the new SPHERE/IRDIS and SINFONI observations and data reduction. In Sect.~3 we report the results and analysis of the data. We draw the two most plausible scenarios for the emission we detect in scattered light, and we compare the quiescence versus outburst near-IR spectra. In Sect.~4 we discuss the results and summarize our conclusions. 

\section{Observations and data reduction} 

\subsection{IRDIS polarimetric observations}

EX Lup was observed with SPHERE (Spectro-Polarimetric High-contrast Exoplanet REsearch, \citealt{Beuzit2019}) as part of the ongoing SPHERE guaranteed time program to search for and characterize circumstellar disks. Observations were carried out with the InfraRed Dual-band Imager and Spectrograph (IRDIS; \citealt{Dohlen2008}) subsystem in dual-beam polarimetric imaging mode (DPI, \citealt{deBoer2020, vanHolstein2020}) on May 16, 2017. The seeing during the night varied between 1.04$^{\prime\prime}$ and 1.38$^{\prime\prime}$, with airmass between 1.11 and 1.16 and coherence time ($\tau_0$) between 1.7~ms and 2.5~ms. 
IRDIS was used in DPI mode in H band, together with N\_ALC\_YJH\_S coronagraph with Inner Working Angle (IWA) of 0.15$^{\prime\prime}$. Integration time of 64~s was used for the individual frames in order to avoid detector saturation outside the coronagraph edge. 
The polarimetric cycles were taken by switching the half-wave plate (HWP)  angles between $0^\circ$, $45^\circ$, $22.5^\circ$ and $67.5^\circ$.

\begin{figure*}[!ht]
\begin{subfigure}
  \centering
  \includegraphics[angle=0,width=0.33\linewidth]{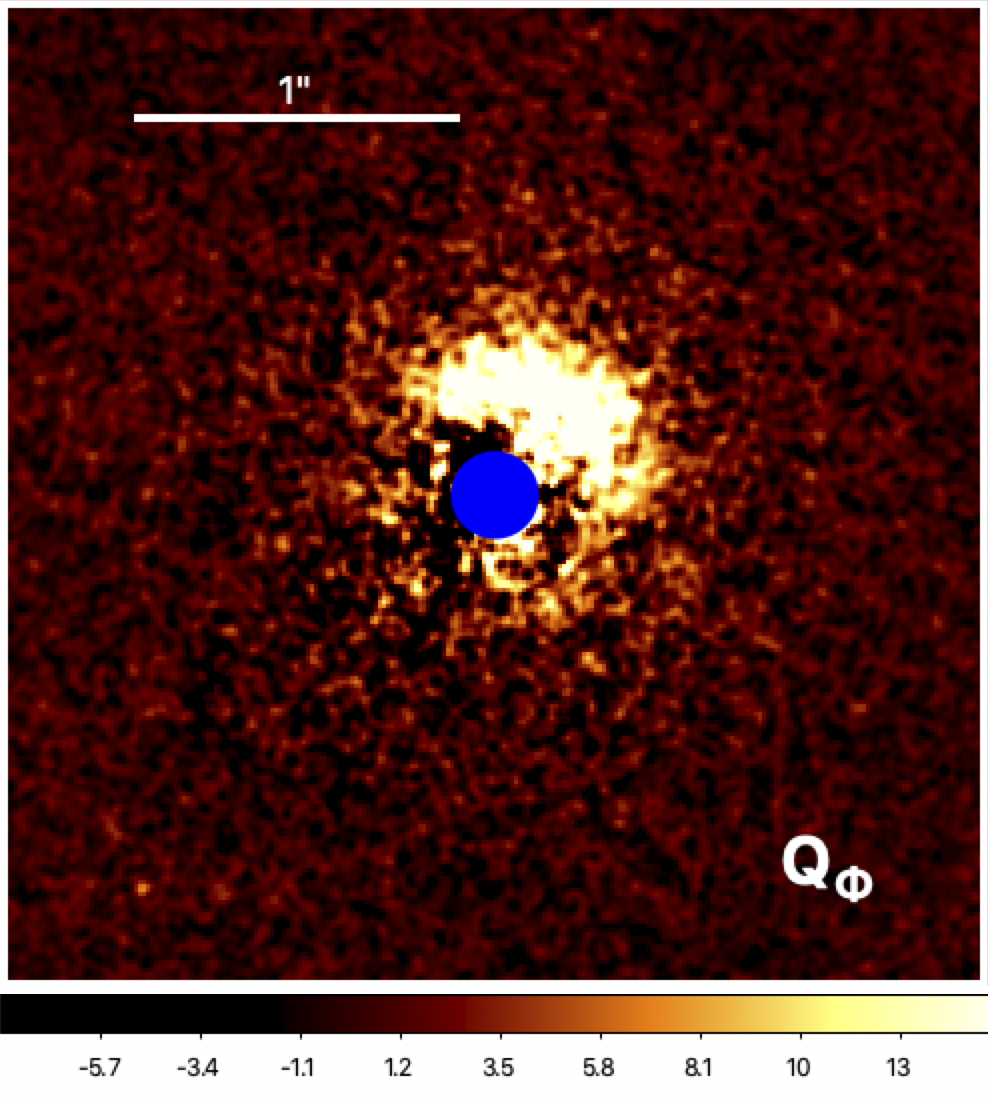}  
\end{subfigure}
\begin{subfigure}
  \centering
  \includegraphics[angle=0,width=0.33\linewidth]{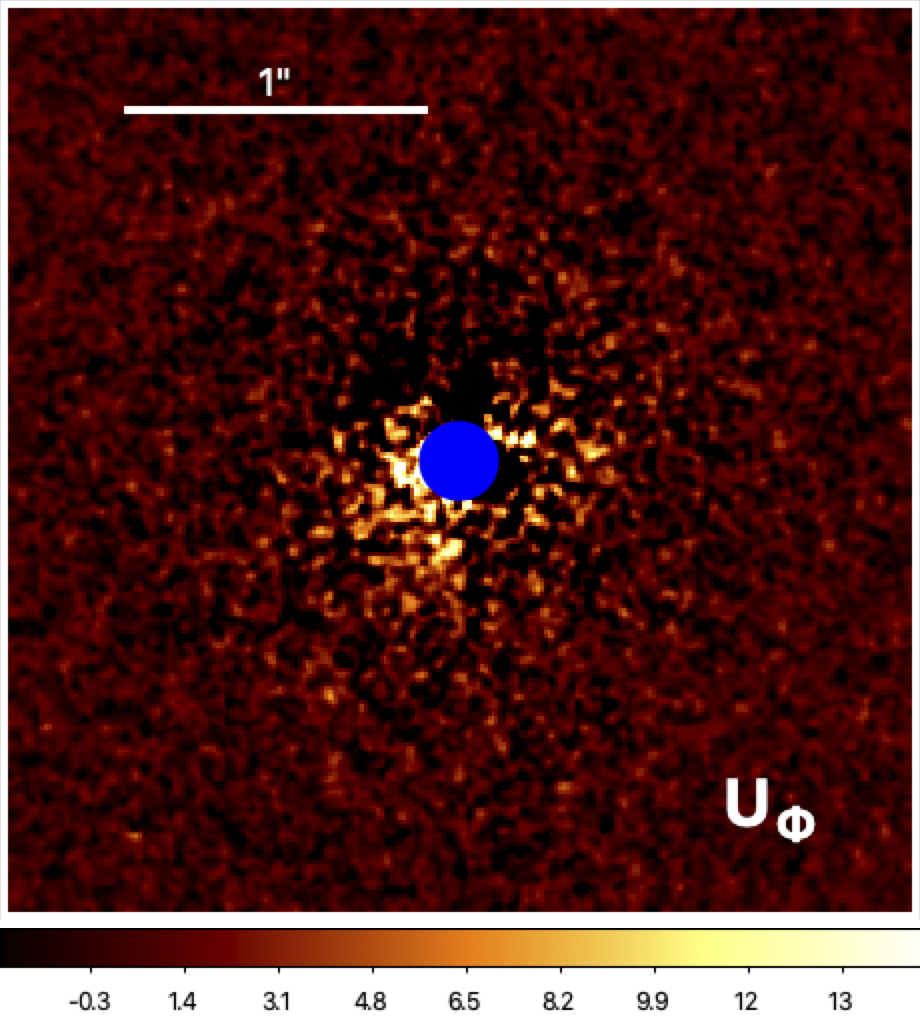}  
\end{subfigure}
\begin{subfigure}
  \centering
  \includegraphics[angle=0,width=0.33\linewidth]{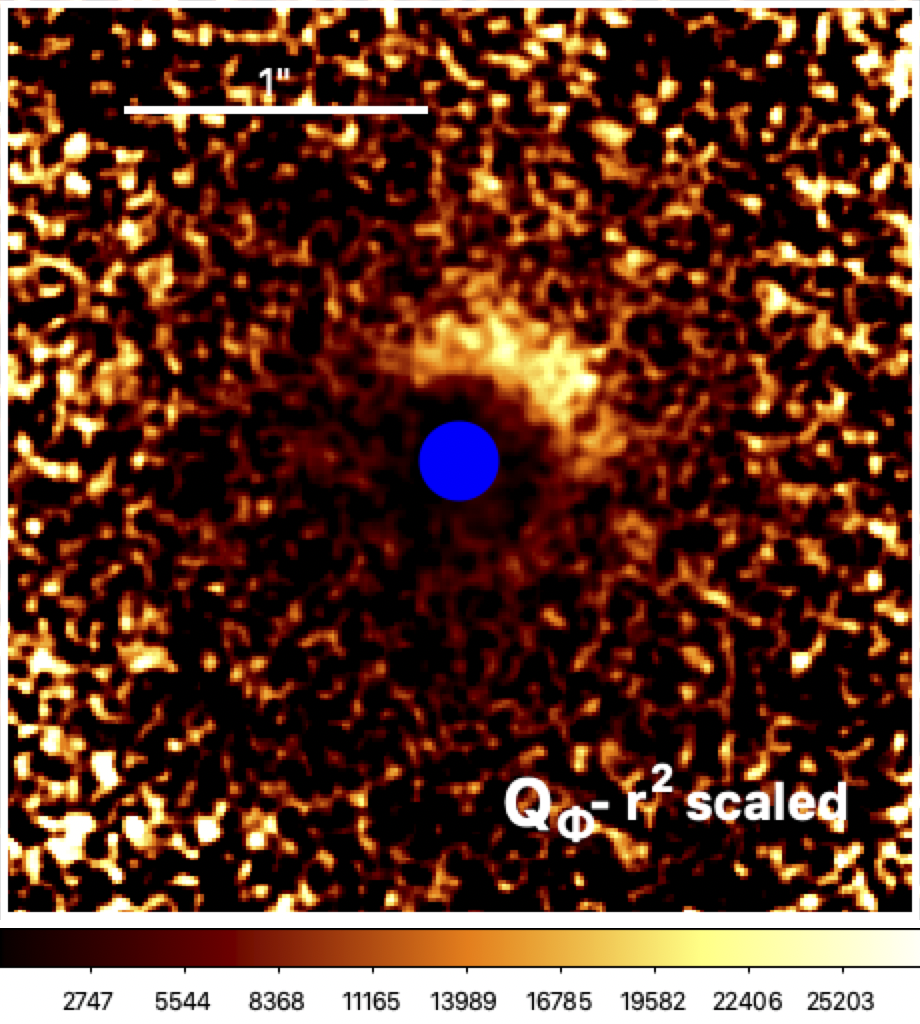}  
\end{subfigure}
\caption{Reduced SPHERE-IRDIS polarimetric Q$_{\Phi}$, U$_{\Phi}$ and Q$_{\Phi}$--r$^2$ scaled images. The coronagraph is marked with a blue circle. We assumed a conservative value for the coronagraph radius of 11 pixels, corresponding to 0.15$^{\prime\prime}$.  
North is up, east to the left.} 
\label{Qphi_Uphi}
\end{figure*}

Data reduction was performed using the IRDAP\footnote{\url{https://irdap.readthedocs.io}} pipeline \citep{van2017, vanHolstein2020}. The pipeline first applies standard calibration to each individual image: dark subtraction, flat fielding and bad-pixel correction. IRDAP then measures the precise position of the central star using the centre calibration frames on the left and right sides of the detector.
Then, the two images in each datacube (NDIT=2) are averaged to obtain one image for each HWP position of the polarimetric cycle.
Finally, the two frames halves of the orthogonal polarization directions are splitted into two individual frames, corresponding to the parallel and perpendicular polarized beam. 
The two orthogonal polarization directions are needed to subtract the non-polarized light coming from the central star, leaving only the polarized light from the circumstellar material. 
To obtain the individual $Q^+$, $Q^-$, $U^+$ and $U^-$ frames for each polarimetric cycle, IRDIS computes the single difference by subtracting the right frame from the left frame. The Stokes Q (and U) are then computed from the double difference using the $Q^+$, $Q^-$, $U^+$ and $U^-$ frames.
Subsequently, IRDAP uses a validated Mueller matrix model describing the telescope and instrument to correct the polarimetric images for instrumental polarization and crosstalk. IRDAP then measures and subtracts the polarization signal of the star from the resulting images.
The subtraction of the stellar polarization signal removes any spurious polarization or polarized astrophysical signal that comes either from the star or from the unresolved part of the circumstellar disk and affects the image of the disk that we want to investigate.  
The Stokes Q (and U) are then obtained by subtracting Q$^-$ (U$^-$) from Q$^+$ (U$^+$). The final Q$_{\Phi}$ and U$_{\Phi}$ images, as shown in Figure~\ref{Qphi_Uphi} are obtained using the formulas given in \citet{deBoer2020}, or using the formulas given in \citealt{Schmid2006b} for Q$_{r}$ and U$_{r}$, considering that Q$_{\Phi}$=-Q$_{r}$ and U$_{\Phi}$=-U$_{r}$. 
The residual signal in U$_{\Phi}$, close to the centre of the image, can be explained either by a suboptimal alignment of the coronagraphic frames, or by multiple scattering in the inner disk.

\subsection{SINFONI observations} 

\begin{figure*}[t]
\centering
\includegraphics[angle=0,width=10cm]{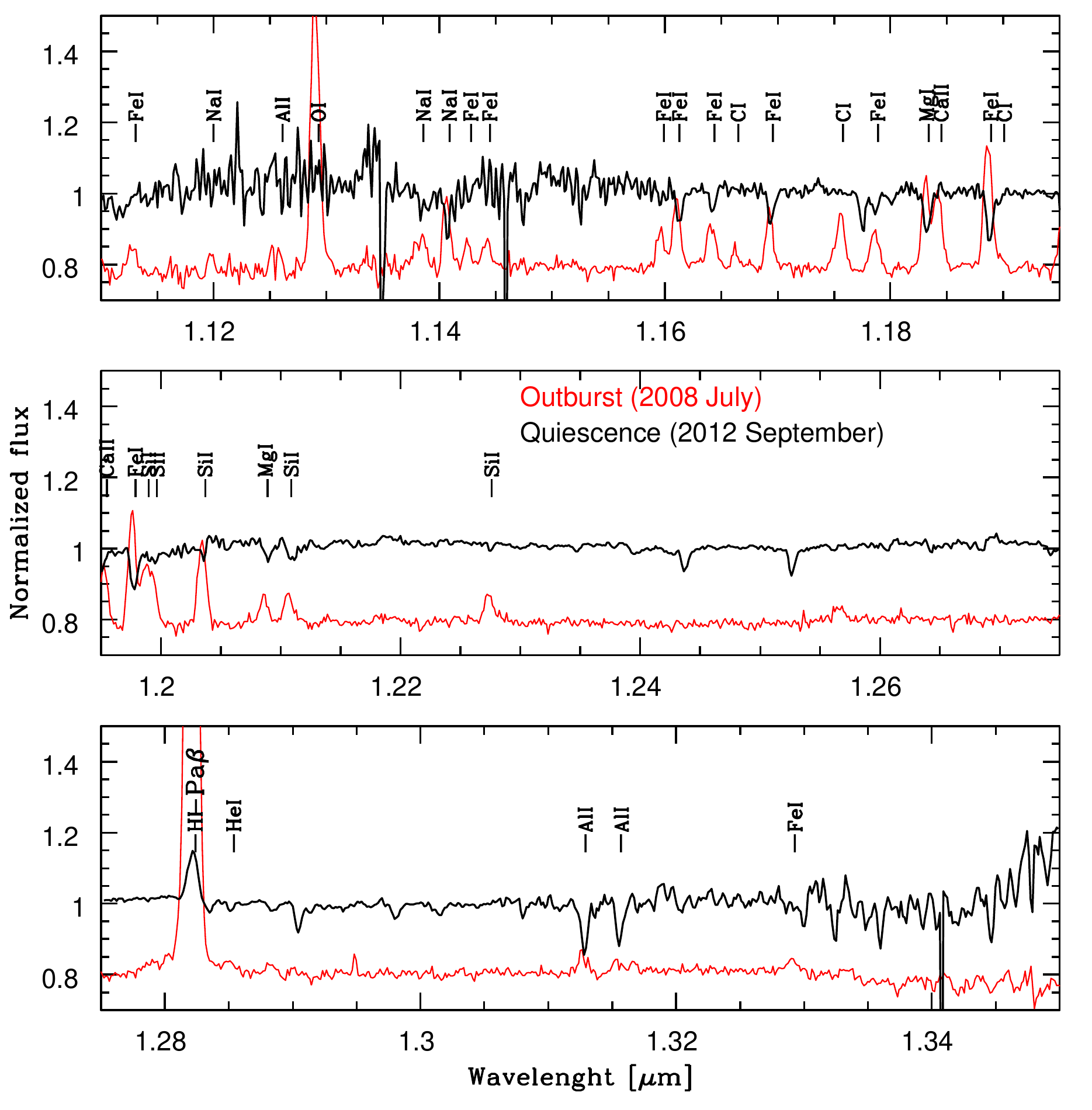}
\caption{Normalized SINFONI J-band spectrum of EX Lup in quiescence compared to the outburst. The position of the emission lines seen in the J-band spectrum taken during the outburst are marked \citep{Kospal2011}. Many lines are seen in absorption rather than in emission, with the exception of the Pa$\beta$ lines, which is still seen in emission. The outburst spectrum is shifted in intensity.}  
\label{all_SINFONI_J}
\end{figure*}

\begin{figure*}[t]
\centering
\includegraphics[angle=0,width=10cm]{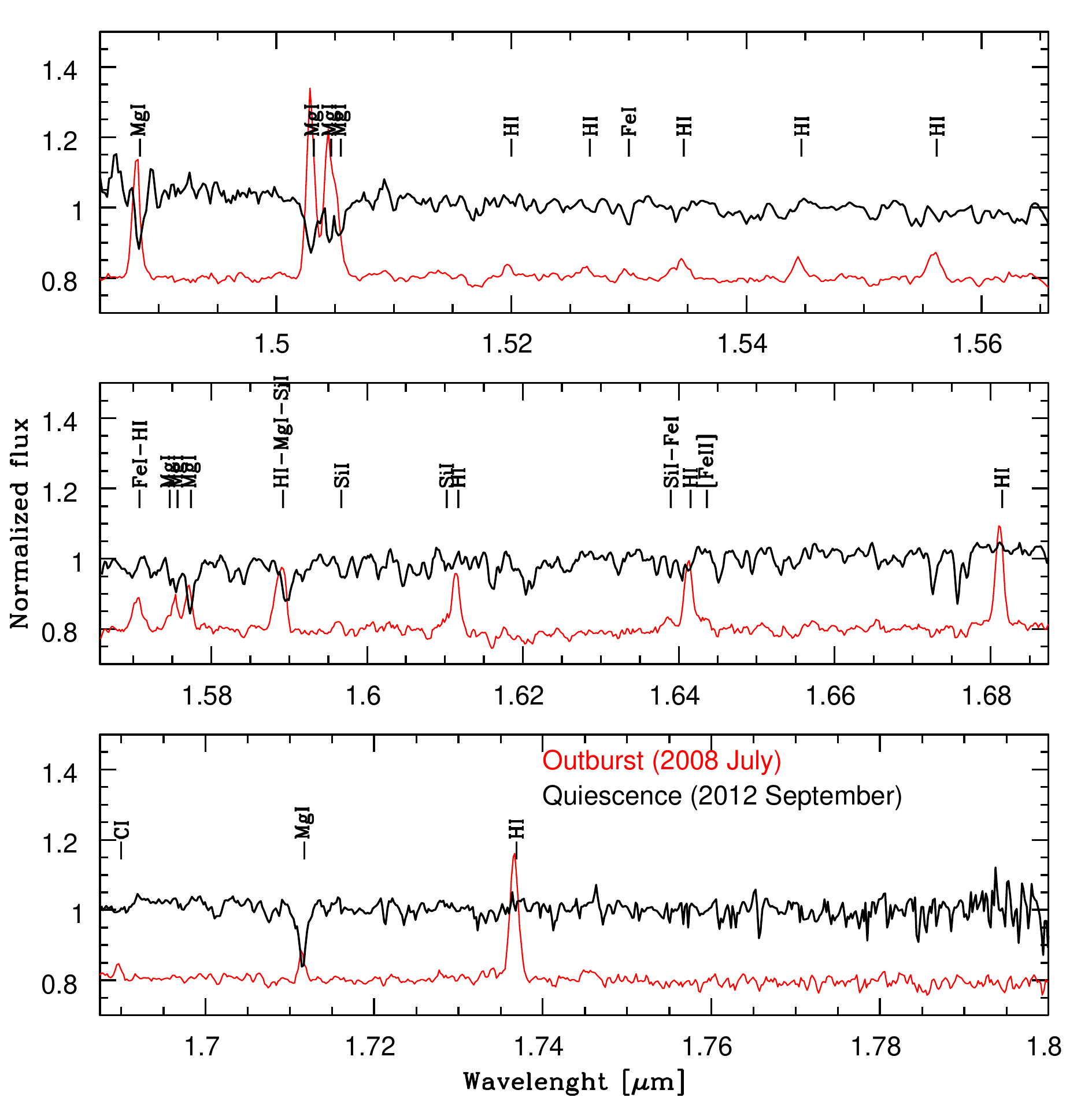}
\caption{Normalized SINFONI H-band spectrum of EX Lup in quiescence compared to the outburst. The position of the emission lines seen in the H-band spectrum taken during the outburst are marked \citep{Kospal2011}. All the lines are seen in absorption rather than in emission. The outburst spectrum is shifted in intensity.} 
\label{all_SINFONI_H}
\end{figure*}

\begin{figure*}[t]
\centering
\includegraphics[angle=0,width=10cm]{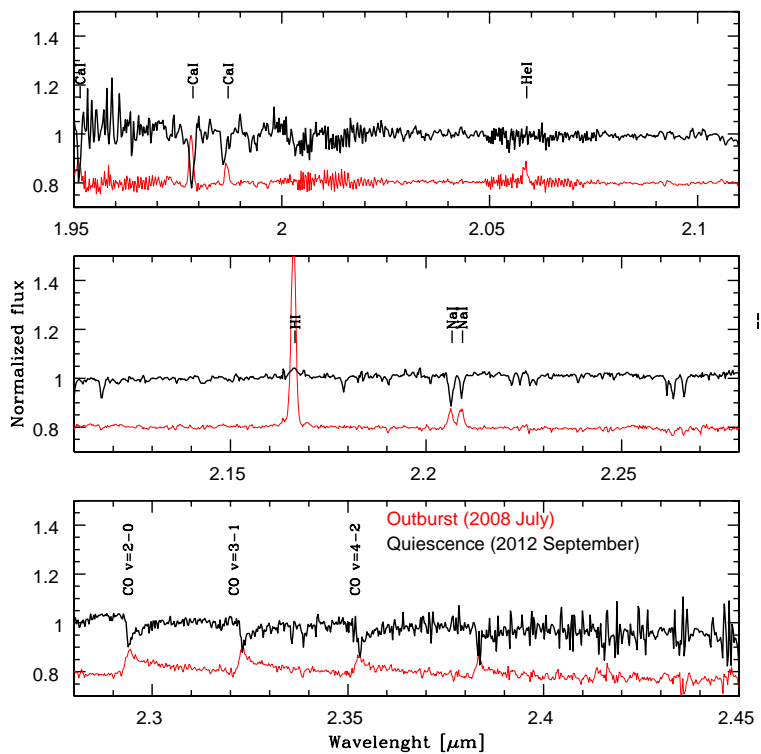}
\caption{Normalized SINFONI K-band spectrum of EX Lup in quiescence compared to the outburst. The position of the emission lines seen in the K-band spectrum taken during the outburst are marked \citep{Kospal2011}. All the lines are seen in absorption rather than in emission. The CO overtone bandhead are also seen in absorption. The outburst spectrum is shifted in intensity.} 
\label{all_SINFONI_K}
\end{figure*}

Medium-resolution near-infrared imaging and spectroscopy of EX~Lup has been obtained with the adaptive optics assisted integral field spectrograph SINFONI \citep{Eisenhauer2003, Bonnet2004} during the quiescent phase, namely on the nights 1/2 September and 2/3 September 2012. The data belong to the program 089.C-0856(A) (P.I. K\'osp\'al) and cover the J$-$ (1.10$-$1.45$\mu$m), H$-$ (1.45$-$1.85$\mu$m) and K$-$band (1.93$-$2.45$\mu$m) with spectral resolution of about 2000, 3000, and 4000, respectively. The images have a spatial scale of 12.5~mas~pixel$^{-1}$ and a field of view of 0.8$^{\prime\prime}\times$0.8$^{\prime\prime}$. The same kind of data were acquired also during the outburst in 2008 \citep{Kospal2011}. 
The data reduction was performed using the version 3.1.1 of the SINFONI pipeline, which performs dark-current subtraction, flat fielding, sky subtraction and wavelength calibration. 
We first calculated the centroid for each image in the data cubes. Then we extracted spectra using an aperture with a radius of 5 pixels and a sky annulus between 20 and 30 pixels to calculate the flux of the star at each wavelength. Telluric absorption features were corrected fitting synthetic transmission spectra using the {\rm molecfit} tool (\citealt{Smette2015, Kausch2015A&A}), and the results were normalized by second order polynomial in the line-free regions of the spectrum. 
Comparison of the spectra taken in the two different nights do not show any significant difference, and they were hence averaged to increase the signal-to-noise ratio (S/N). The resulting spectra are shown in Figures~\ref{all_SINFONI_J}, \ref{all_SINFONI_H} and \ref{all_SINFONI_K}. 
S/N is as high as 200 in the middle of the atmospheric windows and far from strong telluric bands (e.g., at 1.23–1.27 $\mu$m, or at 2.21–2.28 $\mu$m), 30–80 where there is strong telluric absorption (e.g., around 1.18 $\mu$m or 2.01 $\mu$m), and as low as 20 at the edges of the atmospheric windows (e.g., above 1.34 $\mu$m in the J band or above 2.40 $\mu$m in the K band).

\section{Results and Analysis} 

\subsection{Results}

\begin{figure}[!ht]
\begin{subfigure}
  \centering
  \includegraphics[angle=-90,width=\linewidth]{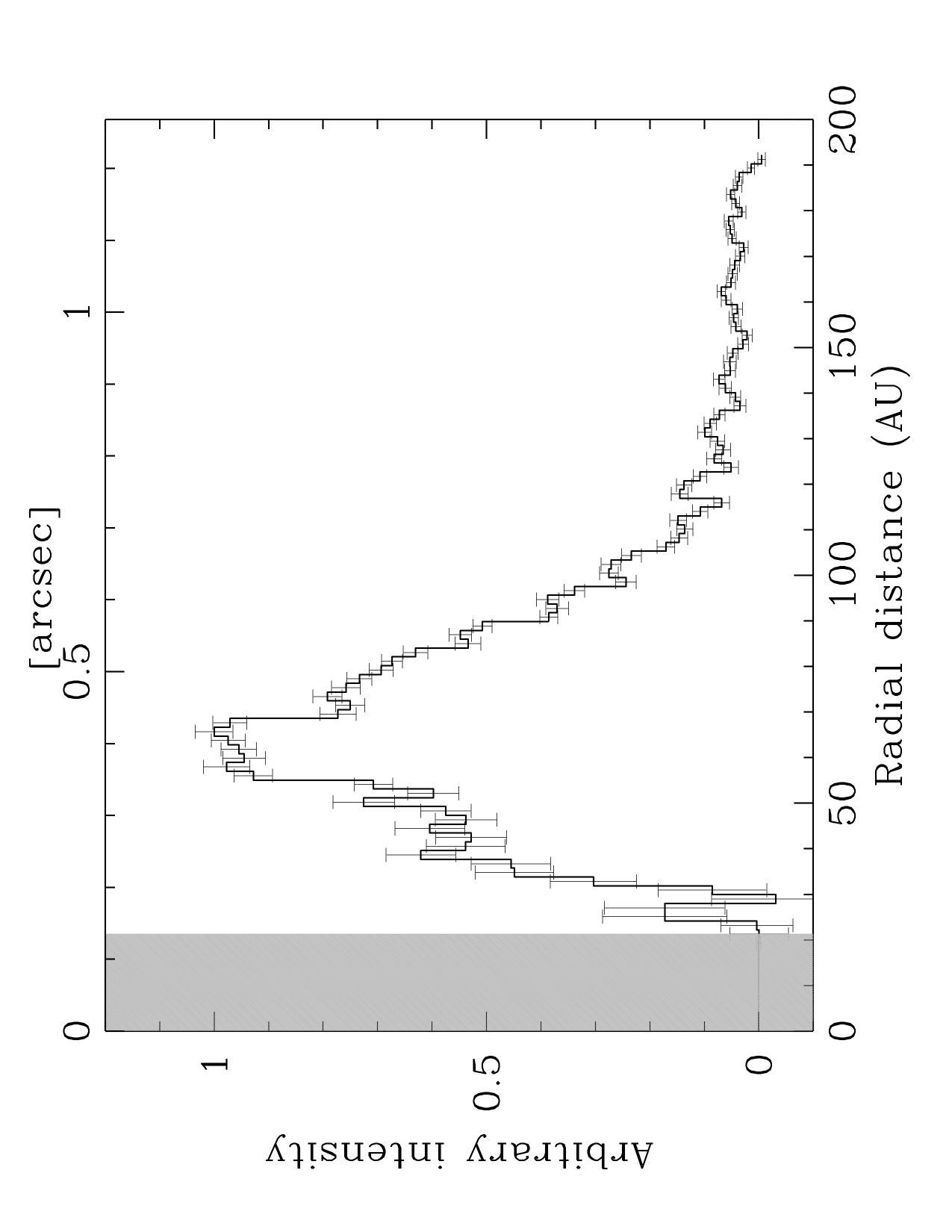}  
\end{subfigure}
\begin{subfigure}
  \centering
  \includegraphics[angle=-90,width=\linewidth]{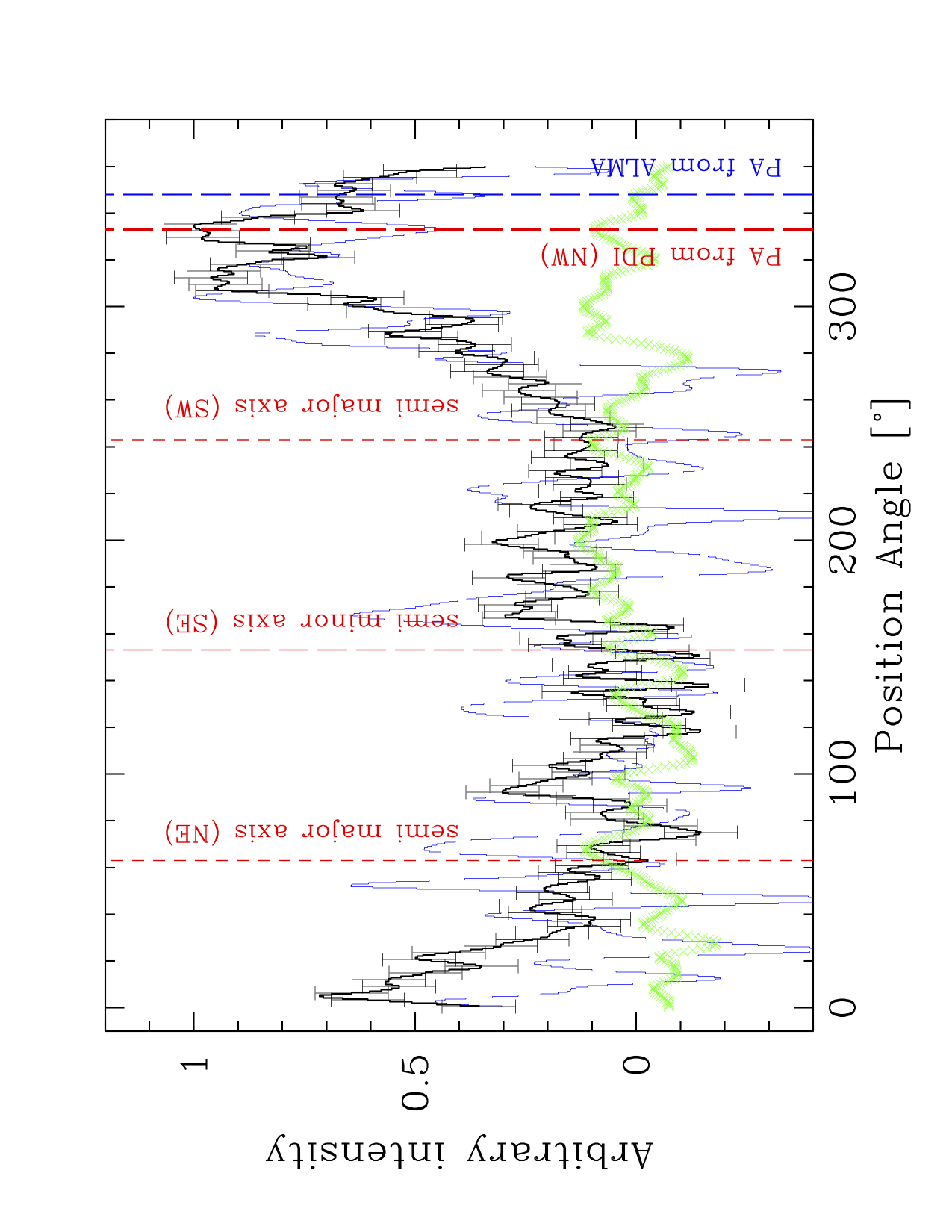}  
\end{subfigure}
\caption{Top panel: radial profile of the H-band deprojected Q$_{\Phi}$ image. The profile is normalized to the brightness peak of the disk. The gray area indicates the radius of the coronagraph. Bottom panel:  the black line shows the azimuthal profile of the Q$_{\Phi}$ image normalized to the unity. The PA of 330$^{\circ}$ as identified from the image is reported as red dashed thicker line. The PA of the outflow as identified from ALMA data is reported as a blue line, and the blue-lighter line show the azimuthal profile as obtained from the $^{12}$CO moment~0 map in the velocity range [-1.6,+1.7], as identified by \citet{Hales2018}. The lighter red lines mark position angles of the semi-major and semi-minor axis, as labeled. The green points mark the azimuthal profile within the $\mu$m-size grains depleted region.} 
\label{profiles_exlup}
\end{figure}

\begin{figure}[!h]
\centering
\includegraphics[angle=0,width=\hsize]{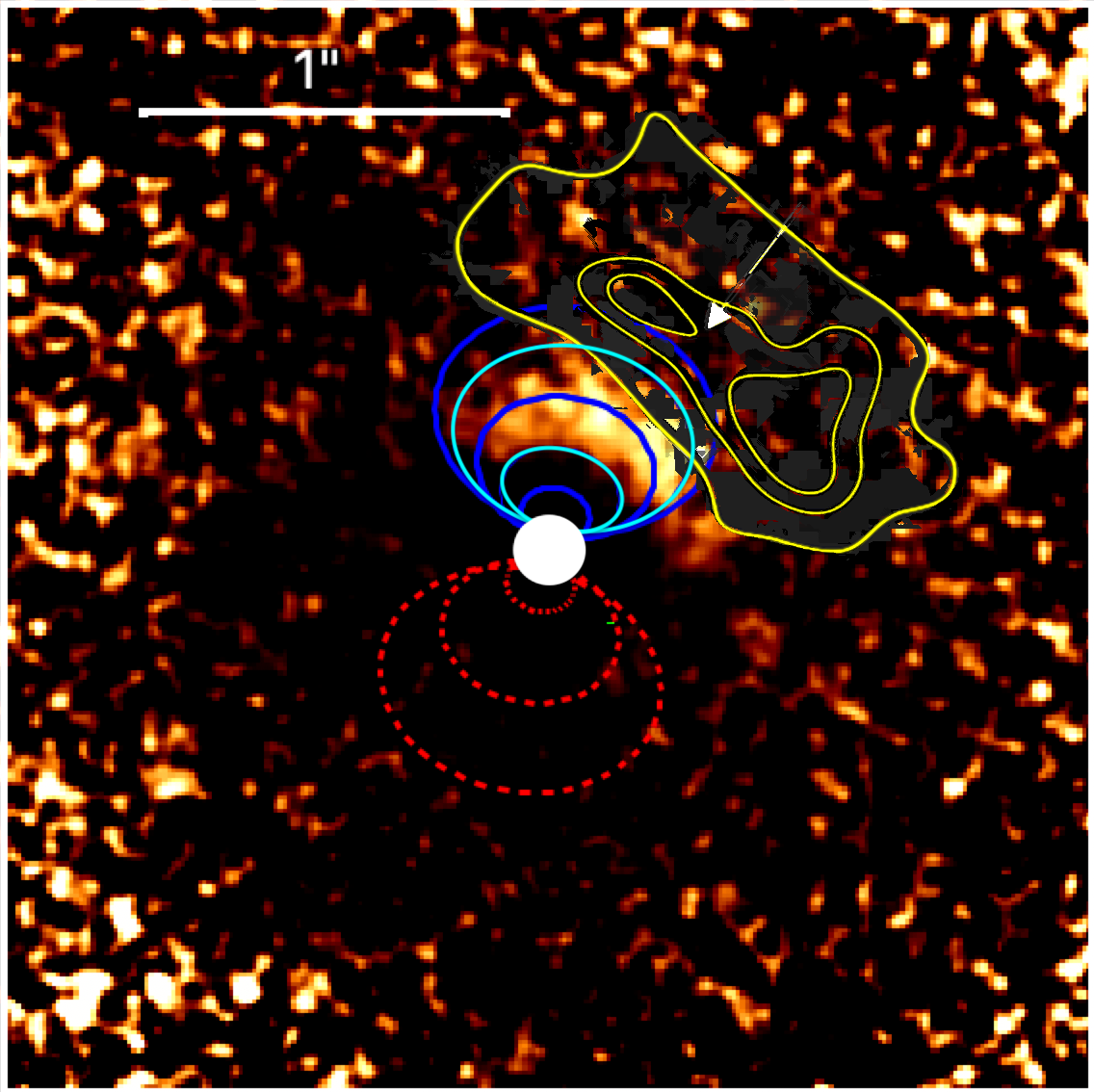}
\caption{Q$_{\Phi}$--r$^2$ scaled image. The blue/cyan and red ellipses highlight the approaching and receding cavity as carved if the emission in scattered light was due to the emission in the cavity walls as for the $^{12}$CO. We report in yellow the extended emission of the $^{12}$CO detected with ALMA.} 
\label{cone_emission}
\end{figure}

The Q$_{\Phi}$ image reported in Fig.~\ref{Qphi_Uphi} shows an increase in the intensity distribution in the North-West direction. In order to account for the  r$^{-2}$ dependency of the stellar flux we show in right panel of Fig.~\ref{Qphi_Uphi} the Q$_{\Phi}$ image multiplied by the square of the separation from the central star, after accounting for the disk inclination. From this image it comes clear that the intensity distribution of the detected feature is asymmetrical, with a banana shape in the North-West side in the SPHERE image in polarized light.
The feature extends radially from $\sim$0.3$^{\prime\prime}$ to $\sim$0.55$^{\prime\prime}$ and is not detected in the South-East side. 

In Fig.~\ref{profiles_exlup}, top panel, we show the radial brightness profile after de-projecting the detected feature, using a position angle of 333$^{\circ}$, as determined through the azimuthal profile (bottom panel of the same figure), and an inclination of 38$^{\circ}$ \citep{Hales2018}. The feature peaks at around 60~au, with a sharp inner edge at $\sim$34~au and a shallower outer edge at $\sim$88~au. There appears to be a plateau in the intensity profile between $\sim$40~au and $\sim$54~au. The region between the coronagraph edge ($\sim$13~au) and the inner edge of the feature appears depleted of $\mu$m-size grains that produce the scattering emission. 
Fig.~\ref{profiles_exlup}, bottom panel, shows the azimuthal profile of the de-projected Q$_{\Phi}$ image. The profile was derived between 34 and 88~au in radial direction. We measured the peak of the intensity and relative error in the azimuthal profile by performing a Monte Carlo approach: we added a normally distributed noise to the extracted azimuthal profile and give as peak value and error the mean and standard deviation of the  distribution of 100 such profiles, respectively. We find for the peak of the azimuthal profile a mean value of 333.1$\pm$0.3$^{\circ}$, and extends between $\sim$280$^{\circ}$ and $\sim$360$^{\circ}$. 
If we assume this is the position angle of the semi-minor axis of the disk detected in scattered light, we can also mark the position on the semi-major axis, as reported in Fig.~\ref{profiles_exlup}. For comparison, we show the 
signal obtained between 23--30~au, within the $\mu$m-size grains depleted region between the coronagraph edge and the inner edge of the feature. 
For comparison, we also show the azimuthal profile of the $^{12}$CO moment~0 map in the velocity range [-1.6,+1.7]. We can notice that, even if the profile is more noisy (we do not include the error bars to improve the readability of the figure), there is an hint of a similar trend of the azimuthal profile between the $^{12}$CO and the scattered light. However, We must also notice that the azimuthal profile has been derived between 34 and 88~au in radial directions, such values' range is optimized for the scattered light image rather than for the $^{12}$CO moment~0 map, which would request larger radial distances as shown in Fig.~\ref{cone_emission}.

\subsection{Analysis}

In the following sections we analyze the results. We compare the near-IR spectra of the outburst versus quiescence phase, then we discuss the emission detected in scattered light exploring two different scenarios. Firstly, we discuss the possibility that the emission is coming from $\mu$m-sized grains in a cavity carved by an outflow. Secondly, we  discuss the possibility that the detected emission is coming from the outer part of the circumstellar disk. 
In order to put both the aforementioned interpretations into context, we summarize and discuss the recent results found by \citet{Hales2018} employing ALMA observation.

\subsection{Outburst vs post-outburst phase in the near-IR}

In Figures~\ref{all_SINFONI_J}, \ref{all_SINFONI_H}, \ref{all_SINFONI_K} we show the SINFONI spectra in J$-$, H$-$ and K$-$band taken in 2012 during the quiescence phase. Contrary to the J, H and K-band spectra of EX~Lup during the outburst, the quiescent-phase spectra do not show the forest of emission lines and is dominated by absorption lines/bands typical of an M-type star. The most prominent emission comes from the Pa$\beta$ line at 1.2821$\mu$m (see Figure~\ref{all_SINFONI_J}), that is however much weaker than observed during the outburst.  The emission features detected by \citealt{Kospal2011} during the burst, and listed in their Table 1, are marked in the figures as vertical tick marks. 
Metallic lines such as Na, Ca, K, Fe, Ti, and Si absorption are numerous in M-type stars, and are often used for spectral type classification \citep{Cushing2005}. Atomic absorption lines such as SI, MgI, AlI, highlighted in Figures ~\ref{all_SINFONI_J}, \ref{all_SINFONI_H}, \ref{all_SINFONI_K}, but also observed by \citealt{Sipos2009} are most probably photospheric in origin (see also \citealt{Herbig2001}). 
The comparison between the spectra taken during the burst and in the post-burst quiescent-phase highlights that there is not ejection activity around the star, as suggested by the absence of the {[Fe\sc{ii}]} lines at 1.25 and 1.64$\mu$m. 
We performed spectro-astrometry of the Pa$\beta$ line in order to check if it might trace extended emitting material moving at different velocities. Pa$\beta$ does not show any spectroastrometric signal, suggesting that it is most likely coming from very close to the star. Spectroastrometric signal in the Pa$\beta$ line was instead found during the burst \citep{Kospal2011} and was likely indicating rotating material around the star.

The absence of emission lines in the SINFONI spectra, with the only exception of the Pa$\beta$ line, suggests that all the lines detected during the burst are related to the accretion or the ejection of material immediately subsequent to the episodic accretion phenomenon.  Pa$\beta$ in the quiescent spectra might still trace accretion, as also found by \citealt{Sipos2009}, but orders of magnitude lower than in the outburst phase. 

\subsubsection{The disk and outflow as seen by ALMA} 

\citet{Hales2018} reported the first spatially resolved observation of the disk around EX~Lup, obtained with ALMA at 0.3$^{\prime\prime}$ resolution in 1.3~mm continuum emission and in the J=2--1 spectral line of $^{12}$CO, $^{13}$CO and C$^{18}$O. 
The compact dust continuum disk does not show indication of clumps or asymmetries in their data, and is consistent with a characteristic radius of $\sim$23~au.  The elliptical Gaussian fitting of the compact continuum emission tracing the disk is consistent with PA of the disk of 63.1$^{\circ}\pm$1.2$^{\circ}$ and disk inclination angle of 32.4$^{\circ}\pm$0.9$^{\circ}$.  
The $^{12}$CO emitting region extends up to $\sim$1.3$^{\prime\prime}$ from the stellar position and is notably more extended than the dust continuum and $^{13}$CO ($\sim$0.5$^{\prime\prime}$) and C$^{18}$O ($\sim$0.3$^{\prime\prime}$) emitting regions. 
The intensity-weighted velocity fields obtained by \citet{Hales2018} for the three CO isotopologues show that the circumstellar gas is rotating. The line emission is consistent with Keplerian rotation at PA=78$^{\circ}$ and disk inclination of 38$^{\circ}\pm$4$^{\circ}$. The gas characteristic radius is 75~au. 
The overall $^{12}$CO emission shows an additional velocity gradient perpendicular to the disk's major axis consistent with the presence of a blueshifted outflow. In order to reconcile the low velocity of the gas ($\sim$2~km/s) with respect to the systemic velocity, they attribute this blue emission to a molecular outflow interacting with the ambient material rather than to gas launched directly from the EXors source during the burst. However they did not detect the redshifted counterpart of the outflow. 
According to their interpretation, the disk's far side is in the North-West and the near side in the South-East with the exposed surface rotating clockwise as seen by the observer, agreeing with the inner disk configuration proposed by \citet{Sicilia2012}.

\subsubsection{Outflow scenario} 

Considering the results obtained by \citet{Hales2018}, we explore here the consistency between the extended emission of the $^{12}$CO seen in ALMA data and the emission seen in polarized light with SPHERE. We must remember that in the first case we are tracing gas, 
and in the latter case we are tracing $\mu$m-sized grains. 
Several models assume that the radial distribution of $\mu$m-sized particles is expected to be similar to the gas distribution, while a large radial separation between the gas and dust is expected for mm-sized particles (e.g., \citealt{Birnstiel2009, Pinilla2012, Pinilla2015}) which may result in dissimilar observed structures at different wavelengths (e.g., HD 135344B - \cite{Garufi2013}, HD~100546 - \cite{Pineda2019}, and PDS70 - \cite{Keppler2019}, among many others).

As shown in Fig~\ref{cone_emission}, the direction of the $^{12}$CO molecular outflow (whose contours are shown in yellow) is roughly consistent with the direction of the emission in polarized scattered light.  If we were to trace the same brightened walls of the cavity excavated by the outflow as for the $^{12}$CO according to \citet{Hales2018}, we would be observing them in scattered light in the right direction. However, there are a few  shortcomings in reconciling the two observations. First of all, the $^{12}$CO, observed in July 2016 is farther away than the scattered light emission as observed in May 2017. This points out that the two images are tracing the same direction, but not the same gas coupled to the $\mu$m-size dust. 
In Fig.~\ref{cone_emission} we also show in blue and red the on-sky projection of the two cones that should represent the approaching and receding sides of the cavity carved into the medium by the outflow, according to the system's geometry summarized in the previous section ({\it i}=38$^{\circ}$ and {\it PA$_{disk}$}=78$^{\circ}$), and assuming the aperture of the scattered light emission is $\sim$80$^{\circ}$  (as extending from $\sim$280$^{\circ}$ up to $\sim$360$^{\circ}$). We can immediately notice that the $\mu$m-size dust detected in polarized light resides in a very confined area of the approaching cone, and none is detected in the receding cone. The system geometry implies that this emission observed in polarized light is coming from a region between $\sim$30~au and $\sim$60~au in the vertical direction, above the disk's plane. These values are given by the two cyan lines indicating the cone surface in Fig.~\ref{cone_emission}.  
If we assume that the $\mu$m-size grains emitting in scattered light were lifted, for example, due to radiation pressure during the burst in 2008, or by a wide-angle wind, we obtain for the observed dust a velocity of $\sim$30~km/s, given as the velocity at which the dust has traveled between 30~au and 60~au in the 9 years between the burst and the SPHERE observations. This velocity is consistent with a low-velocity wind, but it is small if compared to the velocity observed in jets around young stars (e.g., \citealt{Bally2007, Hartigan2011, Nisini2018}). 
The dust in the redshifted side of the outflow is not observed either becasue of an asymmetric/monopolar flow or due to the system's geometry. 
Another possible reason might be that non-symmetric outlfows in young sources can originate at larger radii, as observed for the Class I object BHB07-11 \citep{Alves2017}.

\begin{figure}[!ht]
\begin{subfigure}
  \centering
  \includegraphics[angle=0,width=0.49\linewidth]{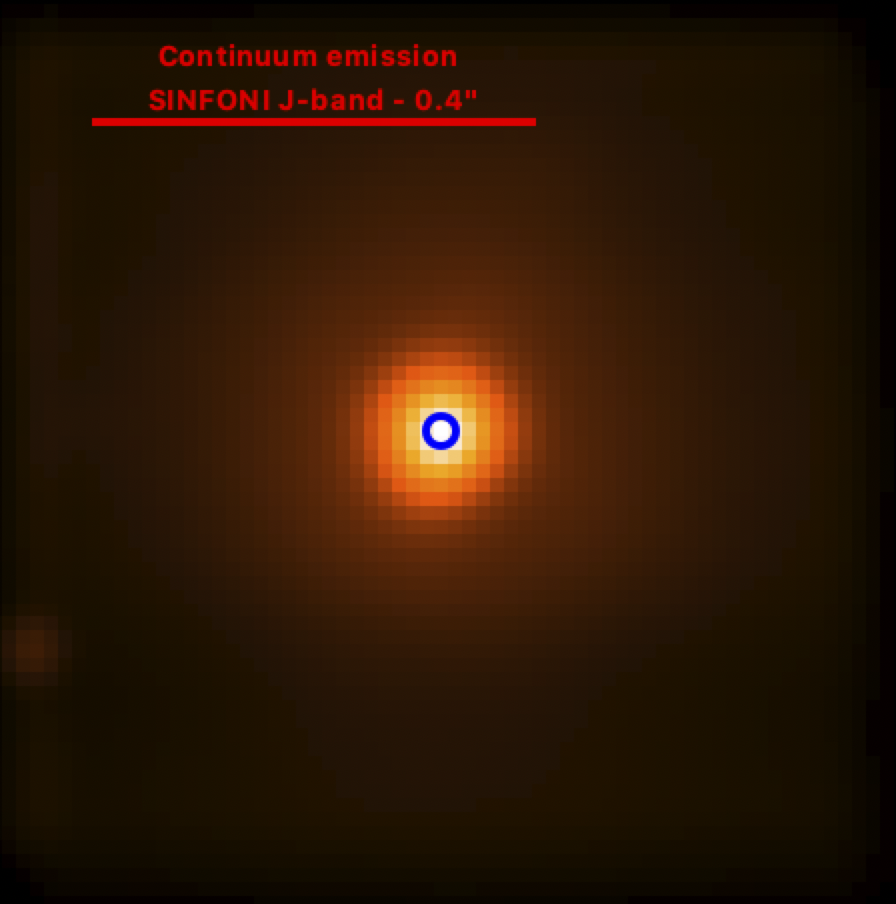}  
\end{subfigure}
\begin{subfigure}
  \centering
  \includegraphics[angle=0,width=0.49\linewidth]{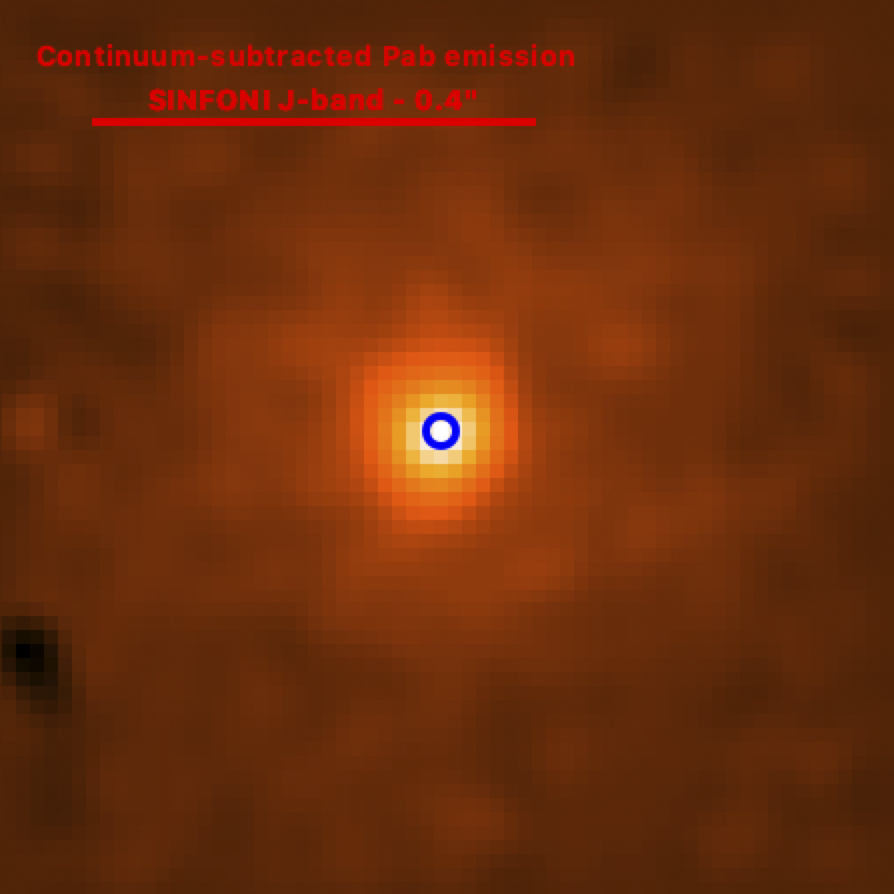}  
\end{subfigure}
\caption{Left panel: SINFONI continuum emission. The blue circle marks the position of the star. Right panel: SINFONI continuum-subtracted Pa$\beta$ emission obtained mediating the channels around the Pa$\beta$ emission line in J-band between 1.281$\mu$m and 1.283$\mu$m. The star position is marked with a blue circle. The emission appears as compact as the continuum emission and does not show evidence of outflowing material. }
\label{sinfoni_pab}
\end{figure}

To find additional elements supporting the outflow scenario we have also looked at the  SINFONI data discussed in Sect. 2.2. The most prominent emission that we see is the Pa$\beta$ emission line. Pa$\beta$ emission is a known tracer of accretion into young T$-$Tauri stars (e.g., \citealt{Muzerolle1998, Natta2002, Alcala2017}) but it has also been proposed as outflow indicator (e.g., \citealt{Whelan2004}). 
In Figure~\ref{sinfoni_pab} we show the 
maps of the Pa$\beta$ line emission and of the adjacent continuum emission from the SINFONI data cube.
The emission in Pa$\beta$ appears centered at the same position as the stellar continuum, without any evidence of elongation or extension along the outflow direction, suggesting that it is consistent with an origin from close to the star, likely accretion onto the star, rather than an outflow or jet. 

\subsubsection{Disk scenario}

In this section we take into account the possibility that the emission in scattered light is coming from the circumstellar disk around EX~Lup. 
As discussed at the beginning of this section, the polarized light emission is asymmetric, and shows a banana-shape, with enhanced intensity in the N-W direction.

\begin{figure*}[!ht]
\begin{subfigure}
  \centering
  \includegraphics[angle=0,width=0.33\linewidth]{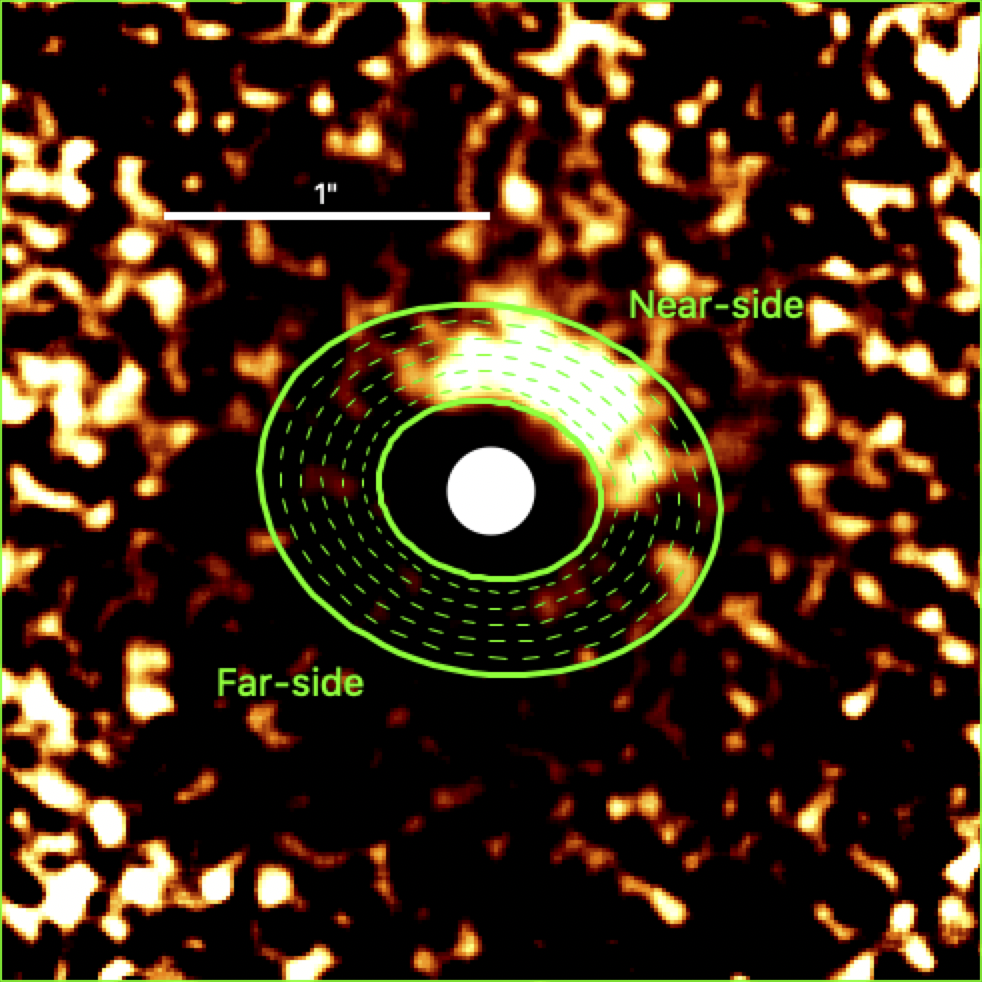}  
\end{subfigure}
\begin{subfigure}
  \centering
  \includegraphics[angle=0,width=0.33\linewidth]{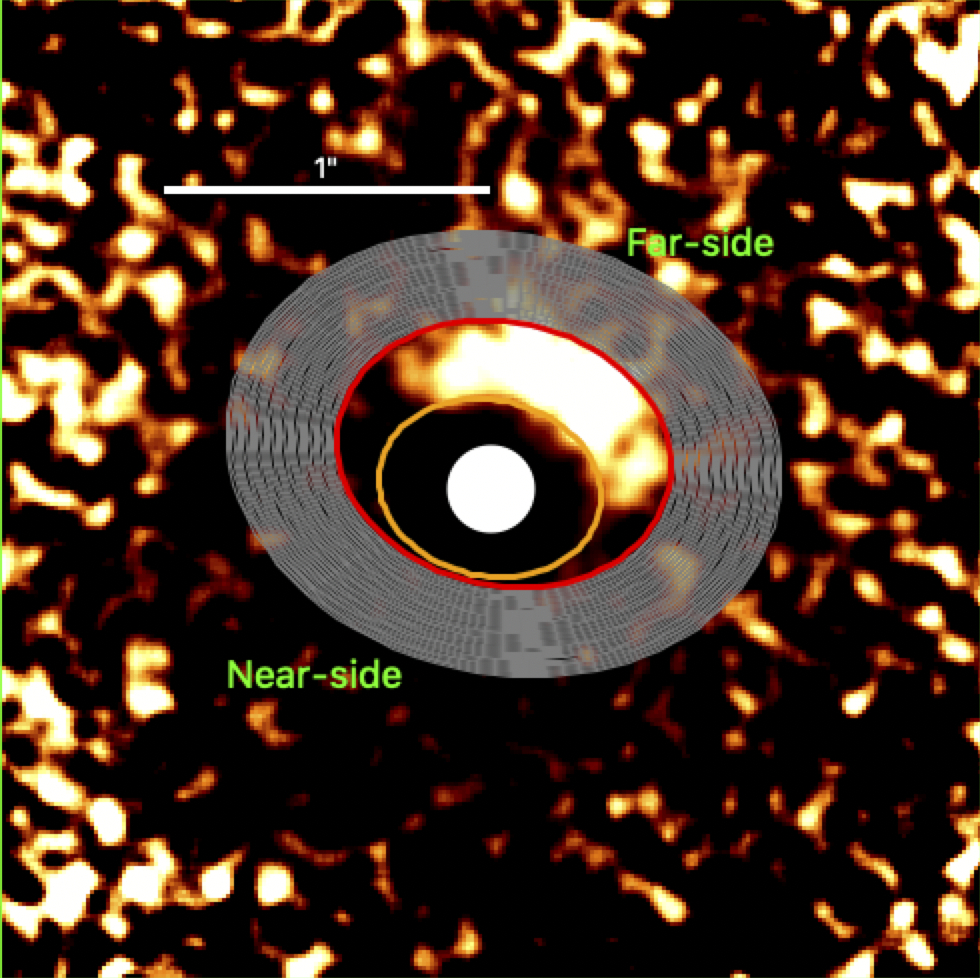}  
\end{subfigure}
\begin{subfigure}
  \centering
  \includegraphics[angle=0,width=0.33\linewidth]{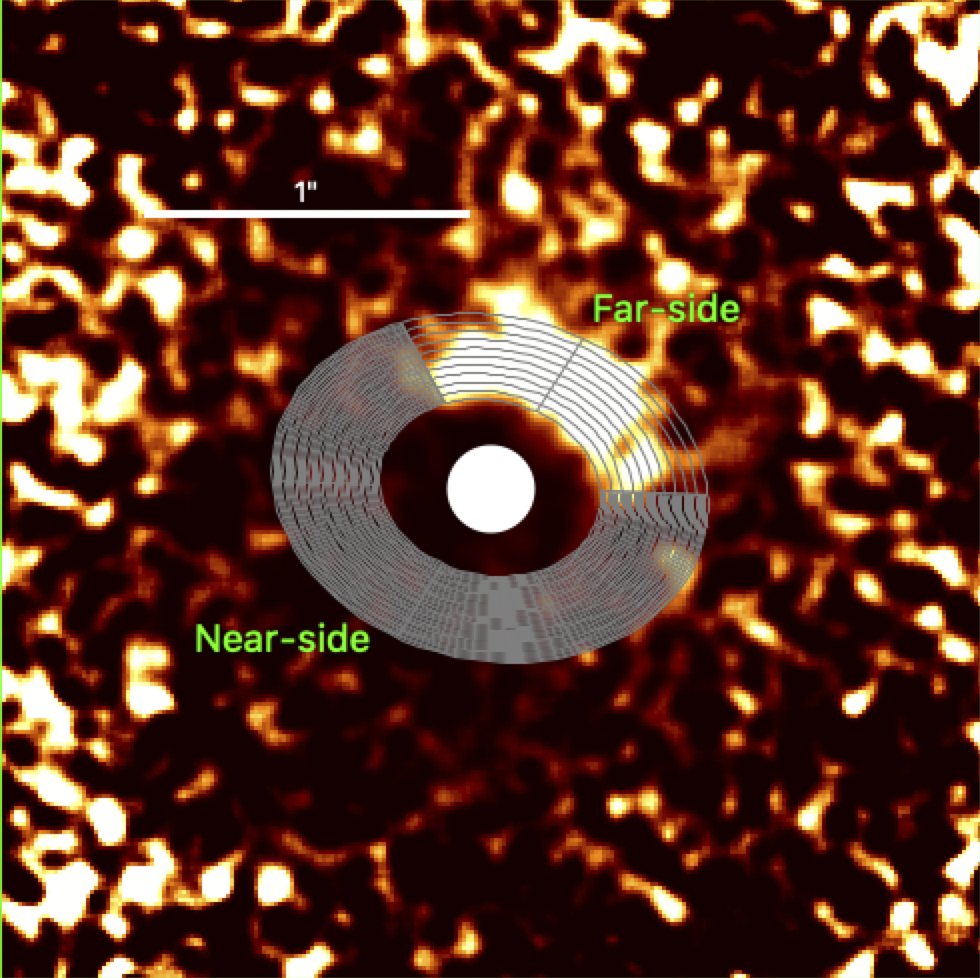}  
\end{subfigure}
\caption{Q$_{\Phi}$--r$^2$ scaled image. Near and far-side of the disk are marked, North is up, East is left. Left panel: the ellipses show the de-projected scattering surface. Near and far-side of the disk are defined according to the brightness asymmetries: brighter side closer to the observer. Middle panel: the brighter side of the disk is coming from the illuminated wall of the outer disk. The orange and red ellipses mark the walls of the disk, the grey-shaded area marks the optically thick disk. Right panel: shadowed disk. The heavy-shaded grey area shows the portion of the disk that is shadowed due to the multiple inner rings misalignment. The light-shaded grey area marks the portion of the disk that is not shadowed, and is emitting in scattered light.  } 
\label{near-far}
\end{figure*}

Brightness asymmetries between the near and far side of the disk are often observed in scattered light from circumstellar disks (e.g., \citealt{Garufi2020} and reference therein) due to the fact that in an inclined disk, the near and far side of the disk surface are seen under a different angle with respect to the star. 
Very small particles scatter light in all directions almost isotropically. When the particle size increases, the scattering becomes increasingly forward peaked \citep{Min2015}. If the dust scatters anisotropically, a brightness asymmetry is observed due to the tendency of photons to be scattered by small angles (an effect yielding a forward-peaking scattering phase function, see, e.g., \citealt{Stolker2016}) with the near side of the disk appearing brighter than the far side of the disk. 
This would suggest that in the EX~Lup the N-W direction represents the side of the disk closer to the observer, while the S-E is the further side, as shown in the right panel of Fig.~\ref{near-far}.  This interpretation is however in contrast with the findings obtained with ALMA by \citealt{Hales2018}, in which the disk is rotating clock-wise with the disk's far side in the N-W. We note that 
polarized light emission above the 1\%$-$2\% level was not detected by \citealt{Kospal2011} in the polarimetric images in H-band filter obtained with NACO in April 2008.  This interpretation is however in contrast with the results coming from ALMA observations.

In order to reconcile the SPHERE data with the system's geometry derived from the ALMA data (in which the brighter side correspond to the disk's further side) we have two different possibilities: either we are detecting scattered light from the inner edge of a ring or the near side of the disk is shadowed, and hence not visible in scattered light. 
In the first case, the scattering surface would be provided by the inner wall of the outer disk on the far side of the star, as for LkCa15 \citep{Thalmann2010}, a K5, $\sim$1M$_{\odot}$ classical T Tauri star with an age of about 3 to 5 Myr \citep{Simon2000}. This interpretation requires a transitional disk with a gap through which we observe the illuminating surface. EX~Lup does show evidence of a dust gap between 23~au (the coronagraph edge) and $\sim$35~au when the sharp edge is seen in the radial profile. Moreover, the disk inclination of EX~Lup ($\sim$38$^{\circ}$) is not very different than the one estimated for LkCa15 (between 44$^{\circ}-$50$^{\circ}$;  \citealt{Thalmann2015, Thalmann2016, Oh2016}). A sketch of this configuration is shown in the middle panel of Fig.~\ref{near-far}, in which the grey area shows the circumstellar disk. As for LkCa15, the illuminated surface of the near-side wall is blocked  by the bulk of the optically thick disk, and the  wall is high enough to cast the outer surface of the disk into shadow, suppressing a forward-scattering signature. 

In the case in which shadows are hindering the detection of the near side of the disk in scattered light, we must assume warps and misalignment between the inner and outer disk. The presence of an inner disk is obvious both from the SED \citep{Sipos2009} and from the accretion onto the star \citep{Sicilia2012}. Narrow shadow lanes (e.g., \citealt{Pinilla2015, Stolker2016, Benisty2017, Casassus2018}), as well as broad extended shadows \citep{Benisty2018} or low-amplitude azimuthal variations (\citealt{Debes2017, Poteet2018}) are observed in scattered light. Of particular interest for this study is the case of HD139614, the young (15.6$\pm$4~Myr, \citealt{Fairlamb2015}) intermediate mass (1.5$\pm$0.1M$_{\odot}$) A7V spectral-type star observed in scattered light by \citealt{MuroArena2020}. HD 139614 shows a very broad shadow region in the outer disk (30–200 au), between approximate position angles 0$^{\circ}$ and 240$^{\circ}$, and the modeling of the disk suggests that the observed asymmetries can be reproduced with a two ring-shaped disk parts that are inclined with respect to each other by $\sim$4$^{\circ}$ and also with respect to the outer disk. 
The case of EX~Lup looks very similar to HD139614, where the shadow cast upon the outer disk around EX~Lup spans about 240 degrees azimuthally. 
Extensive modeling was performed by \citealt{MuroArena2020}, and given the similarity between the two cases, we do not deem necessary similar time consuming computations, and rather we may apply to the case of EX Lup the same conclusion drawn for HD 139614: multiple misaligned disk zones, potentially mimicking a warp with the possibility of a planetary mass companion in the disk, located on an inclined orbit, that would be responsible for such a feature and for the dust-depleted gap responsible for a dip in the SED. A sketch of this configuration is shown in the right panel of Fig~\ref{near-far}. The light-shaded grey area show the portion of the disk that is illuminated, while the heavy-shaded grey area the portion of the disk shadowed. The inner  disk is not shown in this sketch, and falls behind the coronagraph. 

Whether the scattering surface is coming from the inner wall of the outer disk, or by the non-shadowed surface of the circumstellar disk we can make a few consideration that supports the disk scenario. 

Using archival VLTI/MIDI observations, \citet{Abraham2019} derive values for the inner disk inclination and PA of 62$^{+7^{\circ}}_{-20^{\circ}}$ and 81$^{+4^{\circ}}_{-39^{\circ}}$, respectively. These values are in rough agreement, within the large error bars, with the estimates of disk inclination and PA derived by \citet{Hales2018} from ALMA data. Using the spherical law of cosines and the nominal values from VLTI/MIDI and ALMA for these two quantities, we have estimated the misalignment of the inner disk (traced by MIDI) and the outer disk (traced by ALMA), which can span from $\sim$2$^{\circ}$ up to $\sim$40$^{\circ}$. Another important limit to take into account is given by the inner disk flaring with respect to misalignment of the inner and outer disk. In fact, if the disks' misalignment is larger than the inner disk flaring, light from the central star would still illuminate the outer disk in the direction where the PDI image does not show any signal. 
To check this aspect we considered the flaring of the disk as obtained by \citet{Sipos2009} from the modeling of the SED. They retrieve a modest flaring of the disk of $\sim$6$^{\circ}$ semi-aperture angle. If we assume a multiple rings' system, and the inner and outer disk are not much misaligned (less than a few degrees), the configuration of the inner disk casting a shadow on the outer disk is highly plausible. 

Moreover, the comparison between the PA of the disk detected with SPHERE and ALMA points that the scattered light disk has a PA that is more consistent to the compact continuum emission detected with ALMA rather than to the more extended gaseous disk detected in CO. This is in agreement if we assume that the dusty disk is casting a shadow on the outer disk. Indeed, in this case, the PA dependence in scattered light would be dominated by the PA of the directly illuminated disk which is generating the shadow (hence, the dusty disk detected in continuum emission with ALMA).

\section{Discussion \& Conclusions}

The new SPHERE data, acquired in polarimetric differential imaging in H-band, together with SINFONI data in J$-$, H$-$ and K$-$band, have allowed us to investigate the nature of the polarized light emitting region detected around EX~Lup. 
Putting together the literature results on this well studied object with the new observations we discussed different aspects on this object. 

We  compared for the first time the near-IR spectra acquired in J-, H- and K-band with SINFONI in the quiescent-phase with the same data taken during the burst. We find that all the emission features observed during the burst have disappeared in the quiescent period, with the exception of the Pa$\beta$ line which however appears less luminous. We conclude that there are no lines related to ejection activity in the quiescent phase, which supports the idea that the emission in scattered light is coming from a disk rather than an outflow.  

Moreover, we outlined two different scenarios for the origin of the emission seen in scattered light. 
The first one analyzes the possibility that the detected scattered light, due to $\mu$m-size grains, is coming from the brightened walls of a polar cavity excavated by an outflow, as for the $^{12}$CO detected with ALMA. 
This scenario would support the recent finding obtained with ALMA by \citealt{Hales2018}. 
The second one analyzes the possibility that we are detecting a more extended circumstellar disk around EX~Lup. 
In this case, EX~Lup is a system with multiple rings, one causing the accretion, very close to the star, that is behind the coronagraph in the SPHERE images, and  one detected in scattered light extending from $\sim$34~au to $\sim$88~au.  Among this latter scenario we consider the case in which the structure of the disk is similar to LkCa15 \citep{Thalmann2010} and the scattering surface is given by the inner wall of the outer disk, and the case in which the structure of the disk is similar to HD 139614 and the scattering surface is the portion of the disk that is not shadowed due to the misalignment between the inner and the outer disk. 
Even if none of the two scenarios can be totally confirmed or excluded, we favour the latter one, in which the PA of the continuum disk as detected by ALMA is the same as the PA of the scattered light disk as detected with IRDIS-PDI.  
We are aware that ALMA and PDI images are tracing very different regions of the disk, with the first one tracing mainly dust settled in the midplane, and the second one tracing the surface layers of the circumstellar disk. 

In the following we discuss what can be the cause of the depleted region in $\mu$m-size grains that we see between the coronagraph edge and $\sim$30~au. 
Continuum emission at 1.3~mm is detected with ALMA up to 23~au from the central star. On the contrary between 30 and 40~au we notice in the radial profile of the IRDIS-PDI image a steep increase in the intensity, which is often associated with the presence of a disk or ring. 

Photoevaporation due to high-energy radiation from the central star can open gaps and cavities in circumstellar disks (e.g., \citealt{Alexander2014, Ercolano2017}).  \citealt{Ercolano2018} showed that X-ray photoevaporation in transitional disks with modest gas-phase depletion of carbon and oxygen can explain a large diversity of accretion rates and cavity sizes, that can extend up to a few tens of au as the one observed in EX~Lup. Photoevaporation was already invoked by \citealt{Sipos2009} to explain the inner-disk hole detected until 0.2~au, very close to the central star.  
Particles trapping may also be the origin of structures, such as gaps and cavity, in circumstellar disks. The trapping in many cases is due to a massive planet filtering dust particles and trapping mainly the mm-sized particles in the outer pressure bump formed at the edge of the planetary gap (e.g., \citealt{Rice2006, Zhu2011, Pinilla2012} among many others). 
Recently, \citealt{Owen2019} invoked a new mechanism that combines particles trapping and photoevaporation. 
The mechanism focuses on the removal of dust from the pressure trap that is created when photoevaporation opens a cavity in an evolved protoplanetary disk. Radiation pressure can efficiently remove small particles from the surface layers of the disk, in the vicinity of the pressure trap: large dust particles settle towards the mid-plane and the centre of the pressure trap, whereas small particles which are created by the fragmentation of the large particles are lofted above the photosphere. The small particles above the photosphere are driven outwards by radiation pressure and the small particles are then replaced by the collisional fragmentation of larger particles in the mid-plane. The radiation-pressure clear-out of a dusty photoevaporating disk as proposed by \citealt{Owen2019} might simultaneously explain the ALMA and PDI observations.

The presence of a companion around EX~Lup has been largely investigated over the years, in particular to explain the outbursting behaviour of the star. \citet{Kospal2014} proposed the presence of an hypothetical companion to account for the periodic variations in the radial velocities of the photospheric absorption lines seen in the spectra of EX~Lup. According to their studies, the radial velocity (RV) signal could be fitted with a companion of m$sini$ = 14.7M$_{Jup}$ ($\sim$0.014~M$_{\odot}$) assuming an inclination of 38$^{\circ}$ located at 0.06~au from the primary.) in a 7.417~d period eccentric orbit around EX~Lup. \citet{Ghez1997} conducted a multiplicity study in the binary star separation range between 0.1$^{\prime\prime}$--12$^{\prime\prime}$, and detected no companion around EX~Lup in K-band. \citet{Bailey1998} employed spectro-astrometry to detect the signal of a companion around EX~Lup with an accuracy of 70 mas if the spectra of the two stars were very similar, or if one is very much fainter. Also in this case, no signal was detected.  
If we interpret the depleted area in the IRDIS-PDI observations acquired for EX~Lup as carved out by a companion (sub-stellar in mass), this companion would also be the most likely cause of the inner and outer disk misalignment.  

The presence of an innermost companion carving the gap between the rings, that are hidden behind the coronagraph in the IRDIS/SPHERE image, is most likely also the cause of the rings' misalignment. A monitoring of EX~Lup with angular differential imaging observation could certainly help in detecting the possible companion that has been largely suggested in the literature.

\begin{acknowledgements}
E.R. is supported by the European Unions Horizon 2020 research and innovation programme under the Marie Sk{\l}odowska-Curie grant agreement No 664931. 
E.R and S.D acknowledge financial support from the ASI-INAF agreement n.2018-16-HH.0. This work has been supported by the project PRIN INAF 2016 The Cradle of Life - GENESIS-SKA (General Conditions in Early Planetary Systems for the rise of life with SKA) and by the "Progetti Premiali" funding scheme of the Italian Ministry of Education, University, and Research. 
This project has received funding from the European Research Council (ERC) under the European Union's Horizon 2020 research and innovation programme under grant agreement No 716155. T.H. acknowledges support from the European Research Council under the Horizon 2020 Framework Program via the ERC Advanced Grant Origins 83 24 28.
SPHERE is an instrument designed and built by a consortium consisting of IPAG (Grenoble, France), MPIA (Heidelberg, Germany), LAM (Marseille, France), LESIA (Paris, France), Laboratoire Lagrange (Nice, France), INAF Osservatorio Astronomico di Padova (Italy), Observatoire de Geneve (Switzerland), ETH Zurich (Switzerland), NOVA (Netherlands), ONERA (France) and ASTRON (Netherlands) in collaboration with ESO. SPHERE was funded by ESO, with additional contributions from CNRS (France), MPIA (Germany), INAF (Italy), FINES (Switzerland) and NOVA (Netherlands). SPHERE also received funding from the European Commission Sixth and Seventh Framework Programmes as part of the Optical Infrared Coordination Network for Astronomy (OPTICON) under grant number RII3-Ct-2004-001566 for FP6 (2004-2008), grant number 226604 for FP7 (2009-2012), and grant number 312430 for FP7 (2013-2016).
\end{acknowledgements}

\bibliographystyle{aa} 
\bibliography{main.bib} %

\end{document}